\documentclass[10pt,a4paper,twocolumn,english,prb,aps,showpacs,preprintnumbers,amsmath,amssymb]{revtex4}
\usepackage[latin9]{inputenc}
\usepackage{color}
\usepackage{babel}
\usepackage{amsmath}
\usepackage{amssymb}
\usepackage{graphicx}

\makeatletter

\usepackage{dcolumn}
\usepackage{bm}

\makeatother

\begin{document}

\preprint{APS/123-QED}

\title{Superlight small bipolarons from realistic
long-range Coulomb and Fr\"ohlich interactions}

\author{A. S. Alexandrov$^{1,2}$}
\author{J. H. Samson$^{1}$}
\author{G. Sica$^{1,3}$}

\affiliation{$^{1}$ Department of Physics, Loughborough University, Loughborough
LE11 3TU, United Kingdom\\
$^{2}$Instituto de Fisica ``Gleb Wataghin'', Universidade Estadual de Campinas, UNICAMP 13083-970, Campinas, S\~{a}o Paulo, Brasil\\
$^{3}$ Dipartimento di Fisica ``E.R. Caianiello'', Universit\`{a}
degli Studi di Salerno, I-84084 Fisciano (SA), Italy}

\date{\today}

\begin{abstract}
We report  analytical and numerical results on the two-particle
states of the polaronic $t$-$J_{p}$ model derived recently with
realistic Coulomb and electron-phonon (Fr\"ohlich) interactions in
doped polar insulators. Eigenstates and eigenvalues  are calculated
for two different geometries. Our results show that the ground state
is a bipolaronic singlet, made up of two polarons.
The bipolaron size increases with increasing ratio of the polaron hopping integral $t$ to the exchange interaction $J_p$ but remains small compared to the system size in the whole range $0\leq t/J_p\leq1$. Furthermore, the model exhibits a phase transition to a superconducting state with a critical temperature well in excess of
$100$K. In the range $t/J_p\leq1$, there are distinct charge and spin
gaps opening in the density of states, specific heat, and magnetic susceptibility  well above $T_c$.
\end{abstract}

\pacs{71.38.-k}

\maketitle

With few exceptions \cite{max}, it is widely believed that the conventional
Bardeen-Cooper-Schrieffer (BCS) theory and its intermediate-coupling Eliashberg
extension \cite{eli} do not suffice to explain  high temperature superconductivity.
On the contrary, there is growing understanding that the true origin of
high-temperature superconductivity should be found in a proper combination of the
Coulomb repulsion with a significant electron-phonon interaction  (EPI) \cite{acmp}.
The many-body theory of strongly correlated electrons and phonons was originally
developed with the on-site Hubbard repulsion and the short-range Holstein EPI using analytical strong-coupling expansion \cite{Alexandrov1981} and powerful numerical
techniques \cite{trugman} in the framework of the Hubbard-Holstein
and Holstein-tJ models \cite{bonca}. Also the many-body
Coulomb-Fr\"ohlich model, which takes into account a finite range of
 realistic interactions, was proposed \cite{Alexandrov1996} and studied
analytically \cite{Kornilovitch2002} and numerically
\cite{Hardy2009} showing a rich phase diagram with a polaronic
Fermi-liquid, superconductivity induced by mobile bipolarons and a charge-segregated phase. In these and many other studies \cite{aledev} both interactions were introduced as input parameters not directly related to the material.

Recently it has been shown that, in  highly polarizable ionic
lattices,  the \emph{bare long-range} Coulomb and electron-phonon
interactions almost negate  each other giving rise to a novel
 physics described by
the polaronic $t$-$J_{p}$ model \cite{Alexandrov2011} with a
short-range polaronic spin-exchange $J_p$  of phononic origin,
\begin{equation}\small
\mathcal{H}
\equiv -\sum_{i,j}t_{ij}\delta_{\sigma\sigma^{\prime}}c_{i}^{\dagger}c_{j}
+2\sum_{\textbf{m}\neq\textbf{n}}J_{p}(\textbf{m}-\textbf{n})\left(\textbf{S}_\textbf{m}\cdot\textbf{S}_\textbf{n}+\frac{1}{4}n_{\textbf{m}}n_{\textbf{n}}\right)\;.
\label{eq:tJp_Hamiltonian}
\end{equation}
Here the sum over $\textbf{n}\neq\textbf{m}$ counts each pair once only, $\textbf{S}_\textbf{m}=(1/2)\sum_{\sigma,\sigma^\prime}c^\dagger_{\textbf{m}\sigma}\overrightarrow{\tau}_{\sigma\sigma^\prime}c_{\textbf{m}\sigma^\prime}$ is the spin $1/2$ operator ($\overrightarrow{\tau}$ are the Pauli matrices), $i=(\textbf{m},\sigma)$ and
$j=(\textbf{n},\sigma^{\prime})$ include both site
$(\textbf{m},\textbf{n})$ and spin $(\sigma,\sigma^{\prime})$
indices; $t_{ij}$ is the polaron hopping integral while
$J_{p}(\textbf{m}-\textbf{n})>t$ represents the  exchange interaction
between polarons on different sites from a residual
polaron-multiphonon interaction. It has been proposed that the
$t$-$J_{p}$ Hamiltonian, Eq.(\ref{eq:tJp_Hamiltonian}), has a
high-T$_c$ superconducting ground state protected from clustering
\cite{Alexandrov2011}.

In this work we present numerical and analytical results on the
two-particle eigenstates of the polaronic $t$-$J_{p}$ model as the
building blocks for high-temperature superconductivity.
It is worth noting that there is a wide difference between
\eqref{eq:tJp_Hamiltonian} and the familiar $t$-$J$ model \cite{tJ}
derived from the repulsive Hubbard U Hamiltonian in the limit $U\gg
t$ omitting the so-called three-site hoppings and EPI.
The latter model acts in a projected Hilbert space constrained to no double
occupancy. On the contrary  $t$-$J_{p}$ Hamiltonian,
Eq.(\ref{eq:tJp_Hamiltonian}) has no constraint on the on-site
occupancy since the on-site Coulomb repulsion is negated by the
Fr\"ohlich EPI. The hopping integral $t_{ij}$ leads to the coherent
(bi)polaron band while the antiferromagnetic exchange $J_{p}$ bounds
polarons into superlight inter-site bipolarons. Moreover, the sign
``$+$'' instead of ``$-$'' in the last density-density interaction
term in \eqref{eq:tJp_Hamiltonian} provides an effective repulsion
between pairs preventing their clustering \cite{DetailedPaper}, while the repulsive
$t$-$J$ model favors a phase separation.

Also different from any model proposed so far, all quantities in the
polaronic $t$-$J_{p}$ Hamiltonian \eqref{eq:tJp_Hamiltonian} are
defined through the  material parameters, in
particular $t_{ij}=T(%
\mathbf{m}-\mathbf{n})\exp[-g^2(\mathbf{m}-\mathbf{n})]$ with
\begin{equation}\small
g^2(\mathbf{m})={\frac{2 \pi e^2}{{\kappa \hbar \omega_0
V}}}\sum_\mathbf{q}{1-\cos(\mathbf{q }\cdot \mathbf{m})\over{q^2}}\;,
\label{t}
\end{equation}
and
\begin{equation}\small
J_p(\mathbf{m})=T^2(\mathbf{m})/2g^2(\mathbf{m})\hbar\omega_0\;,\label{J}
\end{equation}
where $\kappa =\epsilon _{\infty }\epsilon _{0}/(\epsilon _{0}-\epsilon _{\infty })$ and $V$ is the normalization volume.
Here the high-frequency, $\epsilon _{\infty }$ and the static,
$\epsilon _{0}$ dielectric constants as well as  the optical phonon
frequency, $\omega_0$ and the bare hopping integrals  in a rigid
lattice, $T(\mathbf{m})$ are  measured and/or found using first-principle Density Functional Theory \cite{dft} in a parent polar insulator.

In the following we restrict  the range of the exchange interaction
and of the hopping to nearest neighbors. One can readily find
\cite{DetailedPaper} highly degenerate two-particle energy levels of the model
\eqref{eq:tJp_Hamiltonian} for $t=0$,
\begin{equation}\small\label{eq:E0_StaticLimit}
E_{0}(t=0)=-J_{p}\;,\;E_{1}(t=0)=0\;,\;E_{2}(t=0)=J_{p}\;.
\end{equation}
The ground and the highest energy states are bipolaronic
spin-singlet and spin-triplet,  respectively, made up of two
polarons on neighboring sites. The zero-energy states are combinations of pairs of polarons separated by more than one lattice parameter and on-site bipolarons,
since there are no on-site interaction terms in the
Hamiltonian \eqref{eq:tJp_Hamiltonian}.

\begin{figure}[tbp]
\includegraphics[width=1.0\columnwidth]{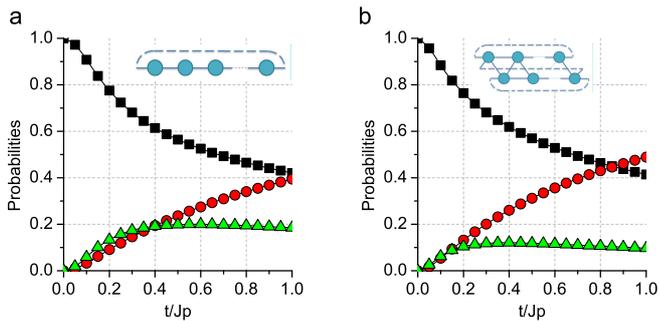}
\caption{(Color online) Probability to find two polarons on the
nearest-neighbor sites: $P_{bp}$ (squares), on more distant sites
$P_{up}$ (circles) and on the same site $D$ (triangles) in the ground state of the $t-J_{p}$ Hamiltonian for chain \emph{(a)} and zig-zag ladder \emph{(b)}.}
\label{fig:Probs_2particle}
\end{figure}

For $t\neq0$  there is a finite bandwidth associated with each of
the three energy levels. Exact diagonalization (ED) results show that the ground
state configuration is virtually unchanged since the distance
between two bound polarons remains of the order of the lattice spacing as
long as $t<J_p$. In that range in fact, regardless of the particular geometry, the probability $P_{bp}$ to find two polarons on nearest neighbor sites decreases gradually with increasing $t/J_p$ and remains finite as shown in Fig.\ref{fig:Probs_2particle}.

\begin{figure}[tbp]
\includegraphics[width=0.7\columnwidth]{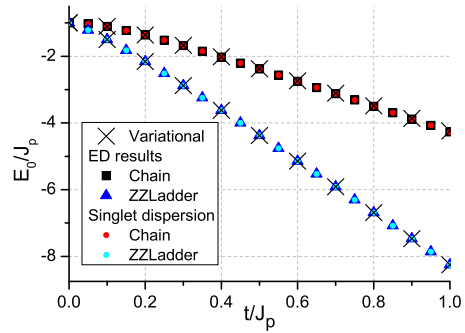}
\caption{(Color online) Two-particle ground-state energy $E_0$ as a function of the hopping. Symbols correspond to ED (squares and triangles) and variational (crosses) data on finite clusters. We also report (circles) the results at $k=0$ obtained by diagonalizing $\hat{H}(k)$ given in Eq.\eqref{eq:MatrixRepresentation}.}\label{fig:E0_2particle}
\end{figure}

We also show in Fig.\ref{fig:E0_2particle} the ground state
energy as a function of the $t/J_p$ ratio for each
analyzed geometry. Importantly, fitting the ED results in the $t\ll J_p$ range, there is a contribution linear in $t$ in the zig-zag
ladder where a single hopping is sufficient for the coherent
propagation of the intersite bipolaron through the lattice
\cite{Alexandrov1996}. On the contrary, in the case of a one-dimensional chain the bipolaron hopping is realized through a
second order process resulting in the quadratic behavior of the
ground state energy as in the case of
the on-site bipolaron \cite{Alexandrov1981}.  As shown in
Fig.\ref{fig:E0_2particle}, ED results are in  excellent agreement with the ones
that can be obtained by means of the variational method
developed by Bon\v{c}a et al. \cite{BoncaMethod}.

Beyond ED results, additional information on the two-particle dynamic can be obtained by considering the following two-particle singlet basis for an infinite lattice:
\begin{equation}\scriptsize
\left|\textbf{m},\textbf{k}\right\rangle=
\left\{
\begin{split}
&\frac{1}{\sqrt{2N}}\sum_{\textbf{n}}e^{i\textbf{k}\cdot(\textbf{n}+\frac{\textbf{m}}{2})}\left(c^\dagger_{\textbf{n}\uparrow}c^\dagger_{\textbf{n}+\textbf{m}\downarrow}+c^\dagger_{\textbf{n}+\textbf{m}\uparrow}c^\dagger_{\textbf{n}\downarrow}\right)\left|0\right\rangle\;,\;m>0\\
&\frac{1}{\sqrt{N}}\sum_{\textbf{n}}e^{i\textbf{k}\cdot\textbf{n}}c^\dagger_{\textbf{n}\uparrow}c^\dagger_{\textbf{n}\downarrow}\left|0\right\rangle\;,\;m=0\\
\end{split}
\right.\;.\label{S=0_basis}
\end{equation}
For the analyzed one dimensional geometries the matrix representation of the $t$-$J_{p}$ Hamiltonian in this basis is:
\begin{equation}\label{eq:MatrixRepresentation}\scriptsize
\hat{H}(k)=\left(
\begin{array}{ccccc}
0               & \sqrt{2}e_1(k)      & \sqrt{2}e_2(k) & 0      & \dots\\
\sqrt{2}e_1(k)  & -J_p+\sqrt{2}e_2(k) & e_1(k)         & e_2(k) & \ddots\\
\sqrt{2}e_2(k)  & e_1(k)              & e_3            & e_1(k) & \ddots\\
0               & e_2(k)              & e_1(k)         & 0      & \ddots\\
\vdots          & \ddots              & \ddots         & \ddots & \ddots\\
\end{array}\right)\;,
\end{equation}
where $e_1(k)=-2t\cos{(ka/2)}$, $e_2(k)=e_3=0$ for the chain, while we have $e_1(k)=-2t\cos{(ka/4)}$, $e_2(k)=-2t\cos{(ka/2)}$, $e_3=-J_p$  for the zig-zag ladder.

The eigenvalues of the tridiagonal matrix $\hat{H}(k)$ determine the energy dispersion $E(k)$. In the  limit $t\rightarrow0$, $e_i(k)\rightarrow0$, so that the ground state
energy is $-J_p$ in agreement with the ED results showed so far. For any $t>0$, the problem is still solvable by requiring the system wave function to decay exponentially in the region where the potential vanishes\cite{DetailedPaper}. In the case of a chain, the energy dispersion can be derived from a cubic equation that, in the $t\ll J_p$ limit, gives:
\begin{equation}\label{E0(k)_chain}\small
  E_s(k)=-J_{p}-(12t^{2}/J_{p})\cos^{2}\left(ka/2\right)+O\left(t^4\right)
\end{equation}
with a quadratic contribution with respect to the hopping term. On the contrary, in the same limit the corresponding dispersion for the zig-zag ladder\cite{CorrectedBipolaronBand} has been found to be linear in $t$:
\begin{equation}\label{E0(k)_zzladder}\small
  E_s(k)=-J_p-t\left[\cos\left(ka/2\right)+\sqrt{1+4\cos^4\left(ka/4\right)}\right]+O\left(t^2\right)
\end{equation}
As shown in Fig.\ref{fig:E0_2particle}, the energy dispersions at $k=0$ obtained for the chain and the zig-zag ladder are in perfect agreement with ED and variational results on finite clusters in the whole range $0\leq t/J_p<1$.

These results allow for some insight   into a possible
superconducting phase transition and pseudogap signatures in
the response functions of the model.

According to the Mermin Wagner theorem \cite{MerminWagnerTheorem},
there should be no phase transition at finite temperatures in 1D and
2D since there is no continuous symmetry breaking. However,  a
finite temperature phase transition in 2D can exist via the
Berezinsky-Kosterlitz-Thouless (BKT) mechanism
\cite{Berezinskii1971}$^,$\cite{Kosterlitz1973}. For a hard-core $2$D
Bose gas, where the Bose-Einstein condensation does not occur
\cite{Gunton1968}$^,$\cite{Hohenberg1967}, a phase transition to a
superfluid state is expected  \cite{Kosterlitz1973}. In particular,
it has been shown \cite{Fisher1988}  in the  dilute limit
$\ln\ln(1/n_{b}r^{2})> 1$ that the critical temperature is
\begin{equation}\label{eq:Tc_superfluid}\small
T_{c}=2\pi\hbar^{2}n_{b}/k_{B}m^{\ast
\ast}\ln\ln(1/n_{b}r^{2}),
\end{equation}
where $n_{b}$ is the boson density per unit area and $r$ is  the range of the boson-boson repulsion. In our case $r\approx a$  and $m^{\ast \ast}$ is the bipolaron mass.

\begin{figure}[tbp]
\includegraphics[width=0.9\columnwidth]{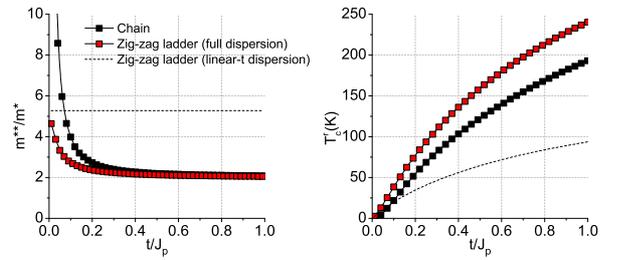}
\caption{(Color online) Ratio of bipolaron to polaron mass
(left panel) in the $t$-$J_p$ model for different lattices and the resulting critical temperature (right panel) estimated with this mass at $n_b=0.01/a^2$ and $J_p=1.0$eV. For the zig-zag ladder we report both the results obtained by using the linear-$t$ dispersion \cite{Alexandrov2011} (dashed line) and the complete one (squares) calculated numerically from \eqref{eq:MatrixRepresentation} by diagonalizing $\hat{H}(k)$.}
\label{fig:mEff&Tc_2Particle}
\end{figure}

To estimate the bipolaron effective mass for a $1$D chain one can use our dispersion \eqref{E0(k)_chain} that gives $m^{\ast\ast}=\hbar^2J_p/6a^2t^2$. In the case of a zig-zag ladder, taking into account only the linear contribution in $t/J_p$ results in an overestimated $m^{\ast\ast}$ with $m^{\ast\ast}/m^\ast\approx5$ where $m^\ast=2\hbar^2/5ta^2$ is the polaron mass \cite{CorrectedBipolaronBand}. Numerical results obtained by solving the eigenvalue problem of \eqref{eq:MatrixRepresentation} show in fact that the ratio $m^{\ast\ast}/m^\ast$ is below this estimate, Fig.\ref{fig:mEff&Tc_2Particle}. In particular, for $t\approx J_p$ we have $m^{\ast\ast}\approx2m^\ast$ for both chain and zig-zag ladder geometries.

Since the effective mass is proportional to $1/t^2$ or $1/t$,
one may conclude that T$_c$ should increase as $t$ or $t^2$ with the polaron
hopping integral. On the other hand, our ED results, Fig.\ref{fig:Probs_2particle},  show that the probability $P_{bp}(t)$ to find a hard-core tightly bound singlet decreases as the hopping increases. At low enough density BEC should not depend on whether the bipolarons are nearest-neighbor or next-nearest neighbor, so long as they are bound and the bipolaron spacing is much greater than the typical polaron separation. On the other hand when bipolarons overlap, their condensation  appears in the form of the Cooper pairs in the momentum space  with a lower critical temperature, rather than in real space (BEC-BCS crossover \cite{Alexandrov1981}). Hence, bounds for the critical temperature can be estimated by weighting Eq.\eqref{eq:Tc_superfluid} with $P=P_{bp}(t/J_p)$ as $T^{r}_{c}\approx P_{bp}(t/J_p)T_c$.
As shown in Fig.\ref{fig:mEff&Tc_2Particle}, despite the  low
carrier density, the critical temperature is about $200$K with the chain and the zig-zag ladder effective mass, for $n_{b}=0.01/a^2$, $a=0.4$nm and $\hbar\omega_0=80$meV.
In particular, $T^{r}_{c}$ obtained for the chain should not be considered as strictly related to the geometry but to the values of $m^{\ast\ast}$ that could be a crude (but quite reliable) estimation of the bipolaron effective mass for a $2$D lattice in the  low-density limit. In the case of cuprate superconductors with  the polaron binding energy in the range $0.5$eV$\leq E_p\leq 1.0$eV and $0.3$eV$\leq J_p\leq1.0$eV $\;$\cite{Alexandrov2011} one  gets the realistic value of  the bare hopping integral $0.2$ eV$\leq T(a)\leq 0.4$eV that gives $0.05\leq t/J_p\leq0.27$ and  the critical temperature $20K\leq T_c\leq 100K$ at $n_b=0.01$.

Let us finally analyse the (pseudo)gap features in the density of
states (DOS) and the spin susceptibility $\chi_s$  of the polaronic
$t$-$J_p$ model at high temperatures well above $T^r_{c}$, when all
carriers are non-degenerate. It is convenient to introduce the
``occupation density of states'' (ODOS), $\rho(\omega,T)$  by
weighting the standard temperature-independent DOS with the
Fermi-Dirac and the Bose-Einstein distribution functions,
\begin{equation}\label{eq:ODOS}\small
\rho(\omega,T)\equiv f_s(\omega,T)\mathcal{N}_{s}(\omega)+2f_{p}(\omega,T)\mathcal{N}_{p}(\omega)\;,
\end{equation}
where:
\begin{equation}\label{eq:DOS}\small
  \mathcal{N}_{s,p}(\omega)=\frac{a}{2\pi}\int_{-\pi/a}^{\pi/a}dk\delta\left(\omega-E_{s,p}(k)\right)\;,
\end{equation}
and $f_{s,p}(\omega,T)=\left[\exp\left((\omega-\mu_{s,p})/k_BT\right)\mp1\right]^{-1}$.
Here $E_{s,p}(k)$ is the (bi)polaron dispersion, $\mu_s=2\mu$, $\mu_p=\mu$ are the chemical potentials of bipolarons and single polarons, respectively, with $\mu<E_s(0)/2$.

\begin{figure}
\includegraphics[width=1.0\columnwidth]{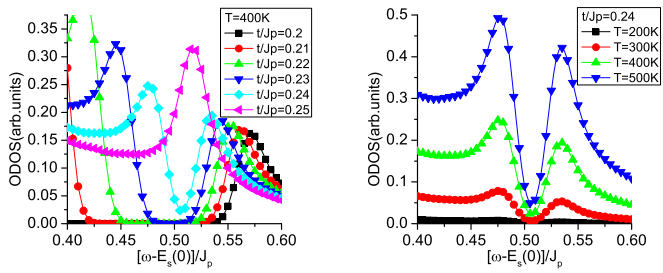}
\caption{\label{fig:DOSsignatures}(Color online) Signatures of  a
pseudogap opening in ODOS for different values of the polaron
hopping (left panel) and temperature (right panel) calculated for the chain with a gaussian broadening $\delta=0.01 J_p$, modeling a disorder effect in the $\delta$-function in Eq.\eqref{eq:DOS}.}
\end{figure}

In the $t=0$ limit, according to \eqref{eq:E0_StaticLimit} we have
three different two-particle energy levels with $E_{s}(k)$ and $E_{p}(k)$ separated by $J_p$. However, at  low carrier density and temperature the highest energy level does not contribute to ODOS and we observe two sharp peaks at $\omega/J_p=-1$ and
$\omega/J_p=0.0$ with a suppression of  ODOS around $\omega/J_p=-0.5$.
Hence there is a single charge/spin pseudogap, $\Delta_c=\Delta_s=J_p$. The ODOS for finite values of $t/J_p$ and temperatures is
shown in Fig.\ref{fig:DOSsignatures}  with some gaussian broadening in the $\delta$-function in Eq.(\ref{eq:DOS}), modeling for instance  disorder effects.
At any $t\neq0$, the two peaks become wider as $t/J_p$ increases and the gap between the bipolaron and the unpaired polaron bands gradually closes. With increasing temperature the  single particle polaron band is  more and more populated along with the incresing population of higher energy levels in the bipolaron band, so that  ODOS reflects a competition between bound and unbound states in the response functions.
In particular, the behavior of the  spin susceptibility indicates the presence of a
finite spin-gap $\Delta_s$ that decreases gradually as $t$ becomes comparable
with the binding energy $J_p$\cite{DetailedPaper}, coherently with the
suppression of the probability to find a bipolaron in the
nearest-neighbor configuration (Fig.\ref{fig:Probs_2particle}).
With increasing temperature,  the number of
occupied states  within the pseudogap also increases.  Without many-body correlation effects (i.e. a screening of the Coulomb and electron-phonon
interactions at finite carrier densities \cite{alebra}), the pseudogap itself does not depend on temperature; it is a matter of whether the temperature is high enough for single polaron states  to have significant occupation. Hence a characteristic
pseudogap temperature $T^{\ast}$ exists in our model above which the pseudogap is suppressed, but  it is a crossover temperature rather than  a critical temperature.
Further signatures of  pseudogap opening are also  found in the specific heat \cite{DetailedPaper} where the Schottky anomaly is induced by thermal excitation within the bipolaronic and the polaronic bands.


In conclusion, we have described some key features of the $t-J_p$
Hamiltonian in the low density limit. We have shown that the ground
state configuration is a small bipolaron singlet. Depending on the
competition between the hopping $t$ and the polaronic exchange
interaction $J_p$, the bipolaron size changes but remains small compared to the system size in the whole range $0\leq t/J_p\leq1$. We have also argued that, in the $2D$
case, the presence of small light bipolarons results in a phase
transition to a superconducting state at a critical temperature in
excess of a hundred K. Finally, the spin susceptibility and the
specific heat of the model revealed a separation of charge and spin
gaps. Because of the presence of a continuum spectrum,
there is no true ground state gap at any finite value of the
polaron hopping $t$. However, strong evidence of a finite pseudogap
has been found in the range where $t<J_p$ above $T_c$.

We gratefully  acknowledge enlightening discussions with  Annette
Bussmann-Holder, Jorge Hirsch, Victor Kabanov, Hugo Keller, Ferdinando Mancini, Frank
Marsiglio, and Roman Micnas,  and the support of the Royal Society
(London) and the UNICAMP visiting professorship program (Campinas, Brasil).



\begin{thebibliography}{99}
\bibitem{CorrectedBipolaronBand}The difference with respect to the one reported in Ref.[11] is due to an incorrect overall sign in the $4$ bipolaron branches. The corrected dispersion gives $50-20\sqrt{5}\approx5$ for the bipolaron to polaron mass ratio, which doubles the estimate of the superconducting critical temperature reported in Ref.[11].
\end{thebibliography}
\end{document}